\documentclass[lettersize,journal]{IEEEtran}
\usepackage{amsmath,amsfonts}
\usepackage{algorithmic}
\usepackage{algorithm}
\usepackage{array}\usepackage{siunitx}							
\usepackage{acro}
\acsetup{single}
\DeclareAcronym{fspl}{
	short = FSPL,
	long = free-space path loss,
}
\DeclareAcronym{psf}{
	short = PSF,
	long = point spread function,
}
\DeclareAcronym{ppb}{
	short = ppb,
	long = parts-per-billion,
}
\DeclareAcronym{ppm}{
	short = ppm,
	long = parts-per-million,
}
\DeclareAcronym{ppt}{
	short = ppt,
	long = parts-per-Trillion,
}
\DeclareAcronym{si}{
	short = SI,
	long = International System of Units,
}
\DeclareAcronym{csi}{
	short = CSI,
	long = channel state information,
}
\DeclareAcronym{rfe}{
	short = RFFE,
	long = radio frequency front-end,
}
\DeclareAcronym{cwtt}{
	short = CWTT,
	long = continuous-wave two-tone,
}
\DeclareAcronym{ipc}{
	short = IPC,
	long = interprocess communication,
}
\DeclareAcronym{awg}{
	short = AWG,
	long = arbitrary waveform generator,
}
\DeclareAcronym{tbp}{
	short = TBP,
	long = time--bandwidth product,
}
\DeclareAcronym{cfo}{
	short = CFO,
	long = carrier frequency offset,
}\DeclareAcronym{sfo}{
	short = SFO,
	long = sampling frequency offset,
}

\DeclareAcronym{fpga}{
	short = FPGA,
	long = field-programmable gate array,
}
\DeclareAcronym{gpu}{
	short = GPU,
	long = graphics processing unit,
}

\DeclareAcronym{nlos}{
	short = NLoS,
	long = non-line-of-sight,
}
\DeclareAcronym{ptt}{
	short = PTT,
	long = pulsed two-tone,
}
\DeclareAcronym{tdma}{
	short = TDMA,
	long = time division multiple access,
}
\DeclareAcronym{tof}{
	short = ToF,
	long = time of flight,
	long-plural = times of flight,
}
\DeclareAcronym{toa}{
	short = ToA,
	long = time of arrival,
	long-plural = times of arrival,
}
\DeclareAcronym{imd}{
	short = IMD,
	long = intermodulation distortion,
}
\DeclareAcronym{fm}{
	short = FM,
	long = frequency modulation,
}
\DeclareAcronym{pm}{
	short = PM,
	long = phase modulation,
}

\DeclareAcronym{xo}{
	short = XO,
	long = crystal oscillator,
}
\DeclareAcronym{tcxo}{
	short = TCXO,
	long = temperature compensated \acl{xo},
}
\DeclareAcronym{ocxo}{
	short = OCXO,
	long = oven controlled \acl{xo},
}

\DeclareAcronym{fir}{
	short = FIR,
	long = finite impulse response
}
\DeclareAcronym{dsp}{
	short = DSP,
	long = digital signal processor
}
\DeclareAcronym{ls}{
	short = LS,
	long = least-squares
}
\DeclareAcronym{qls}{
	short = QLS,
	long = quadratic least-squares
}
\DeclareAcronym{sinc-ls}{
	short = sinc-LS,
	long = sinc nonlinear least-squares
}
\DeclareAcronym{mf-ls}{
	short = MFLS,
	long = matched filter least-squares
}
\DeclareAcronym{rf}{
	short = RF,
	long = radio frequency
}
\DeclareAcronym{lfm}{
	short = LFM,
	long = linear frequency modulation
}
\DeclareAcronym{prf}{
	short = PRF,
	long = pulse repetition frequency
}
\DeclareAcronym{pri}{
	short = PRI,
	long = pulse repetition interval
}
\DeclareAcronym{fmcw}{
	short = FMCW,
	long = frequency modulated continuous-wave
}
\DeclareAcronym{lfmcw}{
	short = LFMCW,
	long = linear frequency modulated continuous-wave
}
\DeclareAcronym{cw}{
	short = CW,
	long = continuous-wave
}
\DeclareAcronym{dbf}{
	short = DBF,
	long = digital beamforming
}
\DeclareAcronym{sar}{
	short = SAR,
	long = synthetic aperture radar
}
\DeclareAcronym{psr}{
	short = PSR,
	long = point scatterer response
}
\DeclareAcronym{rcs}{
	short = RCS,
	long = radar cross-section
}
\DeclareAcronym{crlb}{
	short = CRLB,
	long = Cram\'er-Rao lower bound
}
\DeclareAcronym{dof}{
	short = DoF,
	long = degree of freedom
}
\DeclareAcronym{snr}{
	short = SNR,
	long = signal-to-noise ratio
}
\DeclareAcronym{sinr}{
	short = SINR,
	long = signal-to-interference-plus-noise ratio
}
\DeclareAcronym{fft}{
	short = FFT,
	long = fast Fourier transform,
}
\DeclareAcronym{ft}{
	short = FT,
	long = Fourier transform,
}
\DeclareAcronym{ift}{
	short = IFT,
	long = inverse Fourier transform,
}
\DeclareAcronym{rmse}{
	short = RMSE,
	long = root-mean-square error
}
\DeclareAcronym{psd}{
	short = PSD,
	long = power spectral density
}
\DeclareAcronym{rca}{
	short = RCA,
	long = range of closest approach
}
\DeclareAcronym{rda}{
	short = RDA,
	long = Range-Doppler Algorithm
}
\DeclareAcronym{rma}{
	short = RMA,
	long = Range Migration Algorithm
}
\DeclareAcronym{pfa}{
	short = PFA,
	long = Polar Formatting Algorithm
}
\DeclareAcronym{bpa}{
	short = BPA,
	long = Backprojection Algorithm
}
\DeclareAcronym{rvp}{
	short = RVP,
	long = residual video phase
}
\DeclareAcronym{jrc}{
	short = JRC,
	long = joint radar-communications
}
\DeclareAcronym{doa}{
	short = DOA,
	long = direction of arrival
}
\DeclareAcronym{hci}{
	short = HCI,
	long = human-computer interaction
}
\DeclareAcronym{its}{
	short = ITS,
	long = intelligent transportation systems
}
\DeclareAcronym{rtk}{
	short = RTK,
	long = real-time kinematic
}
\DeclareAcronym{eirp}{
	short = EIRP,
	long = effective isotropic radiated power
}
\DeclareAcronym{gnss}{
	short = GNSS,
	long = global navigation satellite system
}
\DeclareAcronym{imu}{
	short = IMU,
	long = inertial measurement unit
}

\DeclareAcronym{ofdm}{
	short = OFDM,
	long = orthogonal frequency division multiplexing
}
\DeclareAcronym{los}{
	short = LoS,
	long = line of sight
}

\DeclareAcronym{pll}{
	short = PLL,
	long = phase-locked loop
}
\DeclareAcronym{vco}{
	short = VCO,
	long = voltage-controlled oscillator
}
\DeclareAcronym{lna}{
	short = LNA,
	long = low-noise amplifier
}
\DeclareAcronym{if}{
	short = IF,
	long = intermediate frequency,
	short-indefinite = an,
	long-indefinite = an
}
\DeclareAcronym{cots}{
	short = COTS,
	long = commercial off-the-shelf
}
\DeclareAcronym{adc}{
	short = ADC,
	long = analog to digital converter
}
\DeclareAcronym{dac}{
	short = DAC,
	long = digital to analog converter
}
\DeclareAcronym{lo}{
	short = LO,
	long = local oscillator
}
\DeclareAcronym{pcb}{
	short = PCB,
	long = printed circuit board
}
\DeclareAcronym{mimo}{
	short = MIMO,
	long = multiple-input multiple-output
}
\DeclareAcronym{simo}{
	short = SIMO,
	long = single-input multiple-output
}
\DeclareAcronym{mmic}{
	short = MMIC,
	long = monolithic microwave integrated circuit
}
\DeclareAcronym{daq}{
	short = DAQ,
	long = data acquisition
}
\DeclareAcronym{ic}{
	short = IC,
	long = integrated circuit
}
\DeclareAcronym{pa}{
	short = PA,
	long = power amplifier
}

\DeclareAcronym{ti}{
	short = TI,
	long = Texas Instruments
}
\DeclareAcronym{adi}{
	short = ADI,
	long = Analog Devices
}

\DeclareAcronym{roi}{
	short = ROI,
	long = region of interest,
	long-plural-form = regions of interest
}
\DeclareAcronym{v2x}{
	short = V2X,
	long = vehicle-to-everything
}
\DeclareAcronym{av}{
	short = AV,
	long = automated vehicle
}
\DeclareAcronym{cors}{
	short = CORS,
	long = continuously operating reference station
}
\DeclareAcronym{mdot}{
	short = MDOT,
	long = Michigan Department of Transportation
}
\DeclareAcronym{moco}{
	short = MOCO,
	long = motion compensation
}
\DeclareAcronym{sdr}{
	short = SDR,
	long = software-defined radio
}
\DeclareAcronym{gpio}{
	short = GPIO,
	long = general-purpose input/output
}
\DeclareAcronym{usrp}{
	short = USRP,
	long = Universal Software Radio Peripheral
}
\DeclareAcronym{uhd}{
	short = UHD,
	long = \ac{usrp} Hardware Driver
}
\DeclareAcronym{ntp}{
	short = NTP,
	long = network time protocol
}
\DeclareAcronym{ptp}{
	short = PTP,
	long = precision time protocol
}
\DeclareAcronym{lan}{
	short = LAN,
	long = local area network
}
\DeclareAcronym{wlan}{
	short = WLAN,
	long = wireless \ac{lan}
}
\DeclareAcronym{lut}{
	short = LUT,
	long = lookup table
}
\DeclareAcronym{wsn}{
	short = WSN,
	long = wireless sensor network
}
\DeclareAcronym{mac}{
	short = MAC,
	long = media access control
}
\DeclareAcronym{pps}{
	short = PPS,
	long = pulse-per-second
}
\DeclareAcronym{fom}{
	short = FoM,
	long = figure of merit
}
\DeclareAcronym{uwb}{
	short = UWB,
	long = ultra-wideband
}
\DeclareAcronym{twtt}{
	short = TWTT,
	long = two-way time transfer
}
\DeclareAcronym{adas}{
	short = ADAS,
	long = advanced driver assistance systems
}
\DeclareAcronym{swap}{
	short = SWaP,
	long = {size, weight, and power}
}
\DeclareAcronym{cda}{
	short = CDA,
	long = {coherent distributed antenna array}
}
\DeclareAcronym{ap}{
	short = AP,
	long = {access point}
}
\usepackage[caption=false,font=normalsize,labelfont=sf,textfont=sf]{subfig}
\usepackage{textcomp}
\usepackage{stfloats}
\usepackage{url}
\usepackage{verbatim}
\usepackage{graphicx}
\usepackage{cite}
\usepackage{xcolor}
\usepackage{placeins}
\ifcsname hlon\endcsname%
  
\else%
  
\fi%

\input{ieee_submission_notice}

\begin{document}

\title{Fourier Domain Synthesis Imaging Using A Wirelessly Coordinated Distributed Antenna Array}

\author{Sherwin~K.~Shiran~\IEEEmembership{Graduate Student Member,~IEEE},
        Jason~M.~Merlo~\IEEEmembership{Member,~IEEE},\\
        Shivanshu~Ojha~\IEEEmembership{Graduate Student Member,~IEEE},
        Jorge~R.~Colon-Berrios~\IEEEmembership{Graduate Student Member,~IEEE},
        J.~B.~Lancaster,
        and~Jeffrey~A.~Nanzer~\IEEEmembership{Senior Member,~IEEE}%
\thanks{Manuscript received September 00, 2026.}
\thanks{This work was supported in part under the auspices of the U.S. Department of Energy by Lawrence Livermore National Laboratory under Contract DE-AC52-07NA27344 and was supported in part by the LLNL-LDRD Program under Project No. 25-ER-040, in part by the Office of Naval Research under Award N00014-25-1-2208, and in part by the Army Research Laboratory under Cooperative Agreement \#W911NF2420078.}
\thanks{S. K. Shiran, J. M. Merlo, S. Ojha, J. R. Colon-Berrios, and J. A. Nanzer are with the Department of Electrical and Computer Engineering, Michigan State University, East Lansing, MI 48824 USA (e-mail: nanzer@msu.edu).}
\thanks{J. B. Lancaster is with the Lawrence Livermore National Laboratory, Livermore, CA 94550 USA.}
}

\maketitle

\begin{abstract}
In this work we present an experimental demonstration of one-dimensional Fourier-domain imaging using a fully-digital wirelessly coordinated \ac{cda} receiver. The nodes consist of two \acp{sdr} operating with independent system clocks performing wireless time, frequency, and phase coordination without a shared reference such as the \ac{gnss}. Two spatially separated noise sources transmit independent noise waveforms with a bandwidth of \SI{25}{\mega\hertz} and a carrier frequency of \SI{915}{\mega\hertz}. 
%
%
Two sources were positioned at angles of \ang{-6.28} and \ang{12.95} off broadside and imaging was performed.
The two sources are clearly resolved to within \ang{3} of their expected locations in both the individual and combined source measurements and demonstrating the Fourier-domain image reconstruction principle using a fully-digitally coordinated distributed antenna array.

\end{abstract}
\acresetall

\begin{IEEEkeywords}
Coherent distributed antenna arrays, correlation interferometry,
Fourier-domain imaging, software-defined radio, spatial frequency sampling, wireless synchronization.
\end{IEEEkeywords}

\section{Introduction}
\IEEEPARstart{M}{icrowave} imaging maps the angular positions of targets within a scene, making them useful for remote sensing and security applications \cite{Nanzer2012RemoteSensing, Appleby2007Imaging}. Operating in the microwave and millimeter-wave region of the \ac{rf} spectrum instead of optical---as is used in lidar and traditional cameras---enables improved ability to image through dielectric objects as well as increased detection ranges in precipitation and fog owing to their longer wavelengths. 
However, due to the increased wavelength, the physical aperture size required to resolve targets closely separated in angle also increases making fabrication of high-resolution monolithic apertures difficult in some applications. 
Because a conventional monolithic array places its antennas and connections on a single structure, satisfying increasingly demanding resolution requirements requires a correspondingly larger physical aperture, which may ultimately become impractical to construct and deploy. Because the resolution of any aperture is fundamentally limited by the maximum aperture extents, given a fixed operating wavelength, achieving finer angular resolution requires increasing the aperture or array size~\cite{richards2014fundamentals}. This limitation is especially important for airborne and spaceborne systems where the available aperture is limited by the physical dimensions of the platform. \Acp{cda} offer an alternative architecture by allowing independently controlled antenna elements to be placed on separate, potentially mobile platforms. This allows the physical aperture to be reconfigured according to the sensing requirements, without requiring a single large supporting structure~\cite{Nanzer2017OpenLoop,Coutts2006DistributedAperture,Gottinger2021CoherentRadarNetworks,Diao2022UAVSwarm,Krieger2006MultistaticSAR,Werbunat2026CoherentRadarNetworks}. Due to the inherent sparsity of the \ac{cda}, this architecture applies naturally to Fourier-domain imaging. In this technique, each antenna pair forms a baseline that samples a spatial-frequency component of the observed scene by cross-correlating the measured signals to obtain a visibility measurement; once a sufficient number of visibility measurements have been acquired an image may be formed via an \ac{ift} to reconstruct the scene intensity in the spatial domain producing an image similar to that formed by a traditional camera, but with the benefits of a longer imaging wavelength~\cite{Thompson2017Interferometry,Corbella2004Visibility,Vakalis2018NoiseImaging,Luzano2026Thesis,Vakalis2023NonDestructive5G,ColonBerrios2026FrequencyDiverse}. As such, a \ac{cda} provides both the ability to widen and adapt the aperture based on the required angular resolution needed for a given application at a given time. 
Furthermore, by allowing the \ac{cda} elements to move during the image formation process, many spatial frequencies can be synthesized from sparse arrays using a significantly reduced number of elements than would otherwise be required to perform a single snapshot image formation~\cite{chen2025spacetime,luzano2026distributed}.

Existing Fourier-domain imaging systems have largely been implemented using fixed or cabled receiver architectures to observe intrinsically radiating regions of interest. A prominent example of this is performed in traditional radio astronomy where the imaging array passively observes the naturally generated radiation from celestial objects \cite{Thompson2017Interferometry,collaboration2019first}. In fact, passive two-channel radiometers have demonstrated image formation by repositioning the receiving antennas to synthesize a larger aperture \cite{Laursen1998TwoChannel}. Multichannel systems have implemented the correlation and image reconstruction digitally \cite{RamosPerez2012FullyDigital}. Passive radiation, however, can be extremely weak requiring receivers with high gain, low noise, high bandwidth, and long integration intervals \cite{ColonBerrios2026FrequencyDiverse}. Active incoherent imaging addresses this challenge by transmitting noise signals that reproduce the spatial and temporal incoherence required for Fourier-domain reconstruction. This approach has been used to reconstruct 1-D and 2-D images of targets with reduced receiver sensitivity and bandwidth requirements compared with passive interferometry systems \cite{Vakalis2018NoiseImaging}. Additional active imaging work  has demonstrated a 16-element active incoherent millimeter-wave imaging array and a study of the effects of sparse spatial sampling on the reconstructed image \cite{Vakalis2020SixteenElementArray, Vakalis2020ArraySparsity}. Fundamentally, however, these demonstrations of Fourier-domain imaging rely on fixed apertures in which the receiver remains connected to centralized hardware. \Acp{cda} provide more flexible and scalable apertures by placing the receivers on independent platforms. However, this removes shared \acp{lo}, frequency references, cables, and centralized correlation hardware. The major challenge is that this introduces significantly higher relative timing, frequency, phase drift between antenna elements as well as an added challenge of online antenna element localization to estimate each baseline pair's spatial frequency response that may be changing.  Due to the dynamic nature of these systems, even with highly accurate coordination algorithms to correct for electrical state differences between elements, greater estimation errors can be expected. These errors can corrupt the measured visibilities and  image formation. For that reason, significant research has been devoted to the high-accuracy wireless coordination of \ac{cots} \ac{sdr} nodes 
\cite{Nanzer2017OpenLoop,Nanzer2021DistributedPhasedArrays,Ellison2020CombinedRanging,Mghabghab2022PhaseFrequencySynchronization,Prager2020SubnanosecondSynchronization,Gottinger2021CoherentRadarNetworks,Merlo2023PicosecondSynchronization}.
Recently, a fully-digital, real-time method that jointly estimates and compensates the relative time, frequency, and phase offsets between independent \ac{cots} \acp{sdr} without any shared cables or external references was demonstrated providing picosecond-level timing accuracy, enabling this fully-digital Fourier-domain imaging experiment~\cite{Merlo2026FullyDigitalCalibration}. Prior work demonstrated the feasibility of distributed Fourier-domain imaging using a two-element correlation interferometer and a hybrid coordination architecture consisting of wireless \ac{twtt} and one-way analog wireless frequency transfer by performing fully-wireless measurements of a single interferometer baseline~\cite{Luzano2024DistributedInterferometer} as well as partially wireless imaging using the \ac{twtt} method with cabled frequency transfer~\cite{luzano2026distributed,Luzano2026Thesis}.
However, due to the one-way frequency transfer technique, the imaging was highly sensitive to environmental disturbances.  To our knowledge, Fourier-domain image reconstruction using a realtime fully-wireless coordination architecture has not previously been experimentally demonstrated.

In this work, we present a one-dimensional Fourier-domain imaging experiment using a two-node \ac{cda} receiver wirelessly coordinated through a realtime fully-digital time, frequency, and phase calibration technique \cite{Merlo2026FullyDigitalCalibration}. The nodes consist of two \ac{cots} Ettus X310 \acp{sdr} each with independent reference clocks and without external frequency standard such as the \ac{gnss}. To perform the imaging process, one receiver was held fixed, while the other was moved through a series of positions, producing sixteen baselines from $1.5\lambda$ to $9\lambda$ in $0.5\lambda$ steps. Two noise sources are placed in the far-field and generate spatiotemporally incoherent \SI{25}{\mega\hertz} bandwidth noise signals at a carrier of \SI{915}{\mega\hertz}. Four unique scenes were captured by time-division multiplexing the signal into four transmit periods: no transmitters, transmitter A, transmitter B, and transmitters A and B. The received signals were then filtered, divided into their corresponding time-division windows, cross-correlated, and coherently averaged to obtain the complex visibility measurement. These visibility measurements across baselines and are combined via a \ac{ift} to reconstruct the scene.

\section{Fourier-Domain Imaging}
\label{sec:fourier-domain-imaging}

\begin{figure*}[t]
    \centering
    \includegraphics[
        width=\textwidth
    ]{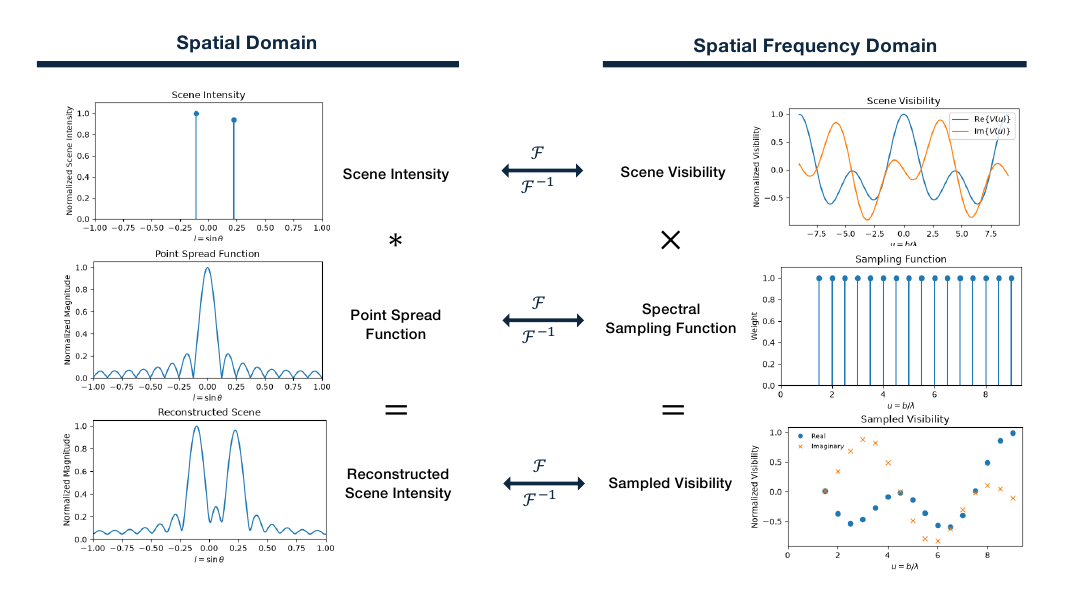}
    \caption{Fourier-domain image formation for a simulated two-target scene.
    The scene intensity and visibility form one Fourier-transform pair, while the baseline sampling function and \acf{psf} form a second Fourier-transform pair. Sixteen positive-baseline samples span $1.5\lambda$ to $9\lambda$ with an increment of $0.5\lambda$. Applying the \ac{ift} to the sampled visibility produces the reconstructed angular scene. }
    \label{fig:imaging_process}
\end{figure*}

The Fourier-domain imaging process, summarized in Fig.~\ref{fig:imaging_process}, can be described using two domains: the spatial domain and the spatial frequency domain. For the one dimensional geometry considered in this work, we express the angular position using the direction cosine: $l=\sin{\theta}$ where $\theta$ is measured relative to broadside. The corresponding spatial frequency coordinate is denoted as $u=D/\lambda$ where $D$ is the physical baseline distance and $\lambda$ is the wavelength. 

Inside the spatial domain the scene is represented by the scene intensity, $I(l)$. The scene intensity describes the time averaged received intensity from each angular direction \cite{luzano2026distributed}. In practice, the scene intensity is unknown and is the quantity that is reconstructed through the imaging process. Inside the spatial frequency domain we can represent the scene as the scene visibility, which follows a direct Fourier transform relationship with the scene intensity
\begin{equation}
V(u)=\int_{-1}^{1}I(l)e^{-j2\pi ul}\,\mathrm{d}l 
\label{eq:scene_visibility}
\end{equation}
During implementation because a finite array contains only a limited number of baselines, it cannot measure the full visibility function at every spatial frequency. Instead each receiver pair defines a baseline $D$ and measures the scene visibility at one spatial frequency coordination $u$. After the electrical state errors between the receivers have been corrected the cross-correlation of their received signals provides a sample of the scene visibility. These spatial frequencies measured by the array are therefore described by a sampling function
\begin{equation}
S(u)=\sum_{n=1}^{N_D} \delta\left(u-u_n\right)
\label{eq:sampling_function}
\end{equation}
where $N_D$ is the number of baseline measurements, $u_n=D_n/\lambda$ is the spatial frequency coordinate associated with the $n$th baseline, and $\delta(u-u_n)$ is an impulse representation of that measured spatial frequency. Given a known sampled visibility then the scene visibility is then the product of the sampling function and the scene visibility
\begin{equation}
V_s(u)= S(u)V(u)
\label{eq:sampled_visibility_product}
\end{equation}
Substituting \eqref{eq:sampling_function} into
\eqref{eq:sampled_visibility_product} gives \begin{equation}
V_s(u) = V(u) \sum_{n=1}^{N_D} \delta\left(u-u_n\right)= \sum_{n=1}^{N_D} V(u_n)\delta\left(u-u_n\right)
\label{eq:sampled_visibility}
\end{equation}
While interferometric process is performed mainly in the spatial Fourier domain, taking the \ac{ift} of the system sampling function yields its \acf{psf}, expressed as
\begin{equation}
\mathrm{PSF}(l)= \int_{-\infty}^{\infty} S(u)e^{j2\pi ul}\,\mathrm{d}u
\label{eq:psf_definition}
\end{equation}
The PSF can be thought of as the impulse response of the system to an ideal point target at broadside. With both domains defined the reconstructed scene is obtained by applying the same \ac{ift} to the sampled visibility. 

\section{Coordination Technique}
\label{sec:coordination-technique}

The distributed receiver uses a fully-digital technique to estimate the relative clock offset and drift between two \ac{sdr} nodes operating with separate system clocks, compensating for the resulting timing, frequency, and phase errors~\cite{Merlo2026FullyDigitalCalibration}. Node, $N_0$, is designated as the reference and a periodic \ac{twtt} with node $N_1$ is used to estimate the initial clock offset and its rate of change, i.e., frequency. Together these estimates form a first order time-varying clock model for node $N_1$. The model is then used to schedule the next receive event on a common time base and digitally correct for the additional time-varying phase error of each of the received samples. The corrected signals are then cross-correlated at each receiver baseline to obtain the complex visibility samples used for Fourier-domain image reconstruction. This enables coherent distributed imaging without a shared hardware clock or external frequency reference.

\subsection{Time and Frequency Estimation}

To estimate time and frequency  two-way method is applied where at each epoch $k$ the nodes exchange a known coordination waveform in both directions and use their local transmit and estimated receive time stamps to form two apparent one-way \ac{tof} measurements
\begin{equation}
\bar{\tau}_{\mathrm{tof}}^{(n \rightarrow m)}[k]
=
T_{\mathrm{sys}}^{(m)}\left(t_{\mathrm{rx}}^{(m)}[k]\right)
-
T_{\mathrm{sys}}^{(n)}\left(t_{\mathrm{tx}}^{(n)}[k]\right)
\label{eq:apparent_tof}
\end{equation}
where $\bar{\tau}_{\mathrm{tof}}^{(n\rightarrow m)}[k]$ is the
$k$th apparent one-way \ac{tof} measurement from transmitting
node $N_n$ to receiving node $N_m$ and the terms
$t_{\mathrm{tx}}^{(n)}[k]$ and $t_{\mathrm{rx}}^{(m)}[k]$ are the
true transmit and receive times, while
$T_{\mathrm{sys}}^{(n)}(\cdot)$ and $T_{\mathrm{sys}}^{(m)}(\cdot)$
are functions which represent the local system clock readings at those times; the term 
\textit{apparent} \ac{tof} is used because it includes clock errors and is thus not the true \ac{tof}. Assuming the scene is quasi-static during the two-way exchange, the physical propagation delay is approximately equal in both direction and the clock offset can be estimated as
\begin{equation}
\widehat{T}_{\mathrm{sys}}^{(n,m)}[k]
= \frac{\bar{\tau}_{\mathrm{tof}}^{(n\rightarrow m)}[k]
-\bar{\tau}_{\mathrm{tof}}^{(m\rightarrow n)}[k]
}{2}
\label{eq:relative_clock_offset}
\end{equation}
where $\widehat{T}_{\mathrm{sys}}^{(n,m)}[k]$ is the estimated clock offset between nodes $N_n$ and $N_m$ during the $k$th two-way exchange. The terms $\bar{\tau}_{\mathrm{tof}}^{(n\rightarrow m)}[k]$ and $\bar{\tau}_{\mathrm{tof}}^{(m\rightarrow n)}[k]$ are the apparent one-way \ac{tof} measurements in the forward and reverse directions.

Once the relative clock offset has been estimated at successive coordination epochs, the relative frequency drift can then determined from the rate at which the clock offset changes with time. Thus the frequency drift is estimated as
\begin{equation}
\widehat{\Delta f}_{\mathrm{osc}}^{(n,m)}[k]=\frac{\widehat{T}_{\mathrm{sys}}^{(n,m)}[k] -\widehat{T}_{\mathrm{sys}}^{(n,m)}[k-1]}{\tau_{\mathrm{twtt}}[k]
}
\label{eq:relative_frequency_error}
\end{equation}
where $\tau_{\mathrm{twtt}}[k]$ is the time between coordination epochs $k-1$ and $k$. Overall, $\widehat{\Delta f}_{\mathrm{osc}}^{(n,m)}[k]$ represents the rate at which the clock of node $N_m$ gains or loses time relative to node $N_n$ which is established as the reference.

\begin{figure}
    \centering
    \includegraphics{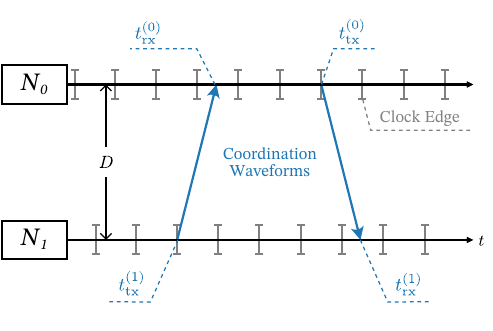}
    \caption{\Acf{twtt} exchange between reference node $N_0$ and node $N_1$. The blue arrows represent the coordination waveforms, and the gray marks indicate the local clock edges. The four labeled events correspond to the transmit and receive times at each node. The full process is explained in~\cite{Merlo2026FullyDigitalCalibration}.}
    \label{fig:twtt}
\end{figure}

The \ac{twtt} process is illustrated in Fig.~\ref{fig:twtt}. Node $N_1$
transmits a coordination waveform to the reference node $N_0$, followed by a transmission in the reverse direction. The local transmit and receive timestamps are used to estimate the relative clock offset as explained before. Repeating this exchange at successive coordination epochs allows the relative rate of clock drift to be estimated.

\subsection{Digital Receive Correction}

Once the predicted clock phase error is derived from the first-order model, the received samples are individually phase-corrected at the digital baseband to account for \ac{cfo}. The corrected samples are given by
\begin{equation}
\widetilde{x}_\mathrm{rx}^{(m)}[i]
= x_\mathrm{rx}^{(m)}[i]\exp\left\{j\left[2\pi \widehat{f}_{\mathrm{cfo}}^{(n,m)}t_s^{(m)}[i] \right]\right\}
\label{eq:receive_phase_compensation}
\end{equation}
where $x_\mathrm{rx}^{(m)}[i]$ and $\widetilde{x}_\mathrm{rx}^{(m)}[i]$ are the uncompensated and
compensated complex received samples at node $N_m$ and $\widehat{f}_{\mathrm{cfo}}^{(n,m)}=f_{0,\mathrm{lo}}^{(m)}\widehat{\Delta f}_{\mathrm{osc}}^{(n,m)}$ is the product of the carrier frequency with the calculated fractional frequency error. The sample times at index $i$ are given by $t_s[i]=i/f_s+t_0$ where $f_s$ is the sampling frequency and $t_0$ is the start time offset of the platform clock. The term $\widehat{\phi}_{\mathrm{start}}^{(n,m)}=2\pi \widehat{f}_{\mathrm{cfo}}^{(n,m)}t_\mathrm{start}$ is a constant phase correction from the predicted initial clock state at the beginning of the receive frame $t_\mathrm{start}$. 

It should be noted that the receive-side correction applied in this experiment compensates for the \ac{cfo} and the phase error predicted from the initial clock offset. In contrast to a single frequency waveform such as a \ac{cw} signal, the noise waveform used here occupies a \SI{25}{\mega\hertz} bandwidth. A timing error therefore introduces a frequency-dependent phase shift across the entire received signal bandwidth rather than only a constant phase offset.  However, while sub-sample time corrections are significant to correct for carrier phase drift, due to significantly larger coherence time of the baseband waveform, the sub-sample time drift has minimal impact on the coherence of the baseband signal over its duration of \SI{10}{\micro\second} in this experiment, as noted in~\cite{Merlo2026FullyDigitalCalibration}. Therefore, the current implementation does not apply a \ac{sfo} correction on the incoming waveforms.

\section{Fourier-domain Imaging Experimental Configuration}
\label{sec:experimental-configuration}

\begin{figure}[t]
    \centering
    \includegraphics{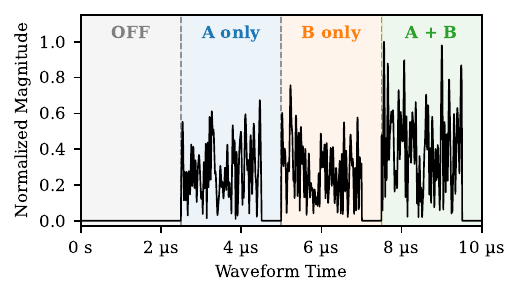}
    \caption{
Normalized magnitude of the transmitted noise waveform. The four states correspond to both sources off, TX-A only, TX-B only, and both sources active. Each active interval contains a \SI{2}{\micro\second} noise waveform followed by a \SI{500}{\nano\second} zero pad, giving a total sequence duration of \SI{10}{\micro\second}.}
    \label{fig:ideal_noise_sequence}
\end{figure}

\begin{figure*}
    \centering
    \includegraphics{
        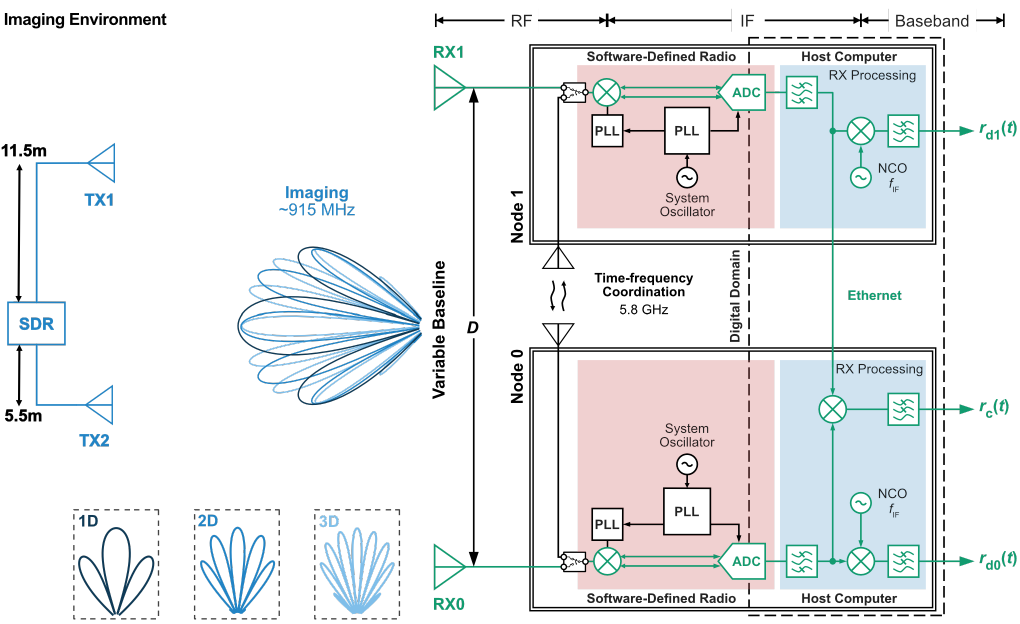
    }
    \caption{Experimental configuration schematic using two distributed receiver nodes and two noise sources. A Ettus X310 \ac{sdr} generates the noise waveforms transmitted by TX1 and TX2. The receiver nodes operate with independent system clocks and are coordinated using the fully-digital technique. Node~0 remains fixed while Node~1 is moved to vary the baseline $D$ from $1.5\lambda$ to $9\lambda$ in increments of $0.5\lambda$, producing sixteen baseline measurements in total.}
    \label{fig:experimental_configuration}
\end{figure*}

The Fourier-domain imaging technique was experimentally demonstrated using a two-node distributed receiver and two spatially separated noise sources. The two receivers operated with independent system clocks and were coordinated using the fully-digital technique described in Section~\ref{sec:coordination-technique}. One receiver remained fixed while the second receiver was moved through sixteen baseline separations. Four separate scenes were imaged by sending a time-division multiplexed on-off pattern of noise pulses from two noise sources as shown in Fig.~\ref{fig:ideal_noise_sequence}. At each separation, the receivers collected the signals produced by the two sources. The full experimental configuration is shown in  Fig.~\ref{fig:experimental_configuration}.

\subsection{Distributed Receiver System}

The distributed receiver consisted of two Ettus Research X310 \acp{sdr}, denoted as $N_0$ and $N_1$, where node $N_0$ was the reference node. Each radio was connected to a Dell OptiPlex 7080 desktop running GNU Radio for wireless coordination and signal processing which were networked together via \SI{1}{\giga bE} for side-channel communication and coordination. The \ac{twtt} coordination exchanges were performed at \SI{5.8}{\giga\hertz} and used two identical 13-dBi Yagi-Uda antennas while the imaging measurements were performed at \SI{915}{\mega\hertz} and used two identical 10-dBi log-periodic antennas. The receivers operated at a samples at a rate of \SI{200}{\mega Sa/\second} on a single channel. An external \acs{rf} switch was used to switch between the coordination and imaging antennas while the \ac{lo} was retuned to each frequency during the switching operation.

\subsection{Noise Waveform}
The imaging signals were generated using two channels on an Ettus X310 \ac{sdr} using an RFNoC replay block. Each transmit channel was connected to a separate 10-dBi log-periodic antenna, producing two independent noise sources, denoted as TX-A and TX-B. The active sources consisted of band-limited pulsed noise waveform with a bandwidth of \SI{25}{\mega\hertz} and duration of \SI{2}{\micro\second}. Independent random seeds were given to each active source so that they could be identified via matched filter in post-processing. 
The ideal four-state noise sequence is shown in Fig.~\ref{fig:ideal_noise_sequence}.

\subsection{Measurement Geometry}

\begin{figure}
	\includegraphics{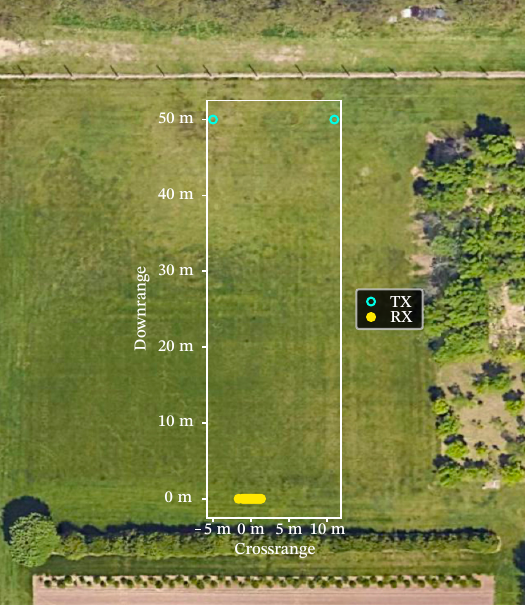}
	\caption{Experiment environment and measurement geometry (Imagery
\copyright 2026 Airbus, Map data \copyright 2026 Google).}
\label{fig:measurement-geometry}
\end{figure}

The experiment geometry is shown in Fig.~\ref{fig:measurement-geometry}. During the measurement, one receiver remained fixed, while the other was moved in $0.5\,\lambda$ steps from $1.5\,\lambda$ to $9\,\lambda$. At the imaging center frequency, \SI{915}{\mega\hertz}, $\lambda\approx\SI{0.328}{\meter}$. Therefore, the physical baseline
separations ranged from approximately \SI{0.491}{\meter} to \SI{2.949}{\meter} in increments of about \SI{0.164}{\meter}. Broadside was defined from the midpoint of the maximum baseline at the $9\,\lambda$. Source A was located
approximately \SI{5.5}{\meter} to the left of broadside, while source B was located approximately \SI{11.5}{\meter} to the right. Both sources
had a downrange distance of  \SI{50}{\meter} from broadside. The corresponding target angles were \ang{-6.28} and \ang{12.95}, respectively.

\subsection{Experimental Simulation}
Before proceeding with the experiment a simulated numerical model was created to determine the expected reconstructed image from the chosen geometry and baselines.
The source locations were defined as $\mathbf{p}_{\mathrm{A}}=(-5.5,50)$\,m and $\mathbf{p}_{\mathrm{B}}=(11.5,50)$\,m relative to broadside reference. The fixed receiver and translated receiver positions replicated the sixteen baseline separations from $1.5\lambda$ to $9\lambda$. The source $s$ and receiver $q$ at the $n$th baseline position gives a calculated propagation distance of
\begin{equation*}
R_{s,q,n}
=
\left\|
\mathbf{p}_s-\mathbf{r}_{q,n}
\right\|
=
\sqrt{
\left(x_s-x_{q,n}\right)^2+
\left(y_s-y_{q,n}\right)^2
}
\label{eq:simulated_range}
\end{equation*}
The corresponding propagation delay was
\begin{equation*}
\tau_{s,q,n}
=
\frac{R_{s,q,n}}{c}
\label{eq:simulated_delay}
\end{equation*}
where $c$ is the speed of light in the medium, and the carrier phase for the different propagation path is
\begin{equation*}
\phi_{s,q,n}
=
\frac{2\pi}{\lambda}
R_{s,q,n}
=
2\pi f_c\tau_{s,q,n}
\label{eq:simulated_phase}
\end{equation*}
For each source and receiver position the transmitted noise waveform
was delayed by the calculated propagation time, rotated by the
corresponding carrier phase, and scaled by the \ac{fspl}. The signal at receiver $q$ is therefore represented as
\[
x_{\mathrm{rx},q,n}[m]
=
\sum_{s\in\{\mathrm{A},\mathrm{B}\}}
a_{s,q,n}
s\left(mT-\tau_{s,q,n}\right)
e^{j\phi_{s,q}}
+
w_{\mathrm{rx},q}
\]
where $s(t)$ is the simulated band-limited noise waveform, $T$ is
the sampling period, $a_{s,q}$ is the \ac{fspl} based amplitude,
$w_{\mathrm{rx},q}$ is receiver noise during the measurement. The noise waveform was modeled as propagated from both
source locations to match the actual experiment. The propagation path loss was modeled as
\[
a_{s,q,n}
=
\left(
\frac{\lambda}{4\pi R_{s,q,n}}
\right)^2
\]
The noise signals had a bandwidth of \SI{25}{\mega\hertz} and a digitally added baseband center frequency of \SI{15}{\mega\hertz}. The receiving \ac{snr} was set to \SI{20}{\decibel}.

To further represent the practical measurement errors: the receiver and transmitter positions had a normally distributed downrange and crossrange position errors added with a standard deviations of \SI{2.5}{\milli\meter} and \SI{5}{\milli\meter}, respectively. In addition, each waveform had a timing error standard deviation of \SI{50}{\pico\second} placed to represent residual coordination error based on previous filed measurements. A total of 2048 Monte Carlo waveform runs were generated at each receiver position. For each run, the signal at the translated receiver was cross-correlated with the signal at the reference receiver. The lag having the largest magnitude was selected and refined via a quadratic least-squares interpolation to refine the estimated delay. The magnitude and phase of this peak were taken to form the complex correlation value for the given receiver separation. After all complex correlation values were obtained from all receiver positions samples were zero padded and an \ac{ift} was applied to simulate the reconstructed image. 

The simulation therefore provided the expected response for
the selected source locations and receiver separations. The red target
markers shown in the simulated image were calculated directly using
\[
\theta_s
=
\tan^{-1}\left(\frac{x_s}{y_s}\right)
\]

\begin{figure}
    \centering
    \includegraphics{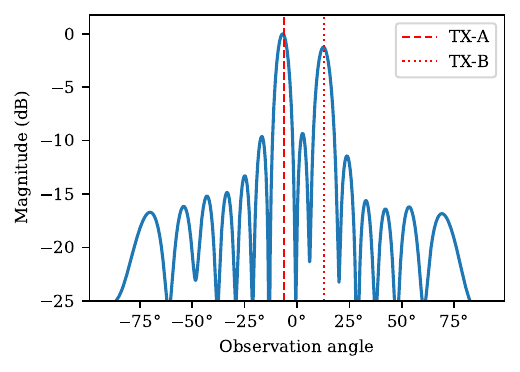}
    \caption{Simulated Fourier-domain image with both noise sources active. The image is reconstructed using sixteen baselines from $1.5\lambda$ to $9\lambda$ in $0.5\lambda$ increments and normalized. The red dashed lines indicate the true source directions.}
    \label{fig:fd_imaging_bothtx_sim}
\end{figure}

The simulated image with both sources active is shown in
Fig.~\ref{fig:fd_imaging_bothtx_sim}, where two separate main
targets are clearly observed at the expected source directions
of $-6.28^\circ$ and $12.95^\circ$.

\section{Experimental Results}
\begin{figure*}
    \centering
    \includegraphics{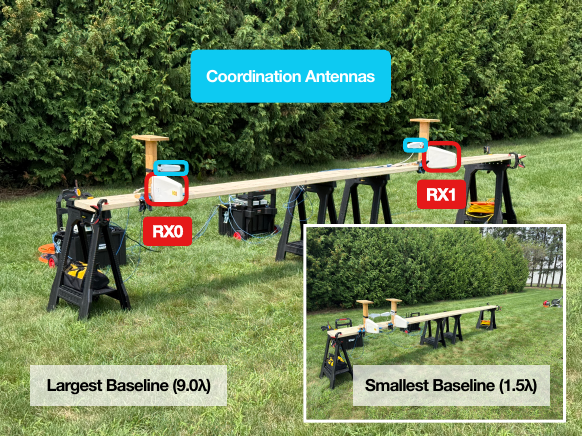}\hfill
    \includegraphics{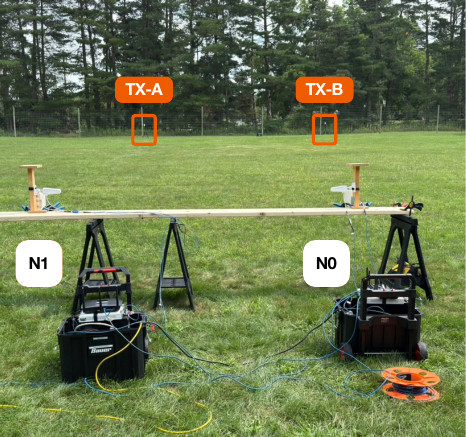}
    \caption{Outdoor experimental setup. Receiver node $N_0$'s antenna RX0 remained fixed while node $N_1$'s antenna RX1 was moved through sixteen baseline separations from $1.5\lambda$ to $9\lambda$ in $0.5\lambda$ increments. Noise sources TX-A and TX-B were positioned approximately \SI{50}{\meter} downrange from broadside and transmitted the noise waveform at \SI{915}{\mega\hertz}.}
    \label{fig:fd_imaging_outdoor}
\end{figure*}

With the numerical model as a reference the Fourier-domain imaging experiment was then experimentally validated in an outdoor environment using the receiver and source geometry described in the previous section. One receiver remained fixed while the second was moved through the sixteen baseline positions, and the two noise sources were placed approximately \SI{50}{\meter} downrange of the receiving aperture. The outdoor experimental setup is shown in Fig.~\ref{fig:fd_imaging_outdoor}.

The recorded receiver signals were processed to obtain a complex visibility measurement for each source state (Source A-only, Source B-only, Source-A and Source -B) and receiver baseline. The received signals were filtered to isolate the transmitted noise signals and minimize out-of-band interference. Following each of the four states for each received capture was sliced to extract each noise source states individually. These corresponding source states were complex cross-correlated and coherently averaged for each baseline. Finally, a \ac{ift} was applied.

\subsection{Received Signals and Filtering}

\begin{figure}
    \centering
    \includegraphics{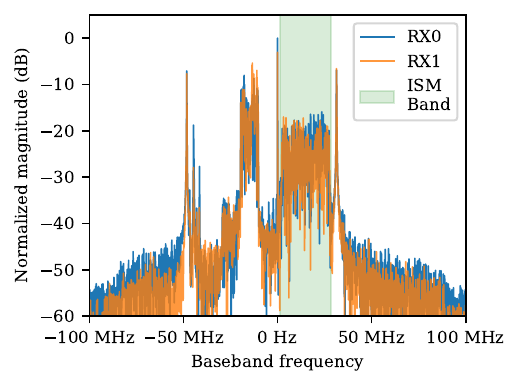}
    \caption{Raw received spectrum for both receivers. The carrier was centered at \SI{900}{\mega\hertz} and a digital baseband modulation of \SI{15}{\mega\hertz} was added to shift the noise signal into the ISM band. The shaded region indicates the desired noise signal band from \SI{2.5}{\mega\hertz} to \SI{27.5}{\mega\hertz}. Both are normalized to the largest \acl{ft} magnitude across the two receivers.}
    \label{fig:raw_receive_spectrum}
\end{figure}

Each receiver capture contained 2000 complex samples at a sampling rate of \SI{200}{\mega Sa/\second}, corresponding to a \SI{10}{\micro\second} receive duration to capture all four states. The receive captures were software triggered and ran at \SI{25}{\hertz}--\SI{30}{\hertz}. Given the baseband noise signal was centered at \SI{15}{\mega\hertz} and had a bandwidth of \SI{25}{\mega\hertz} its frequency range is from \SI{2.5}{\mega\hertz} to \SI{27.5}{\mega\hertz}. Since the received spectrum also contained out of band signal components a frequency domain filter was applied before extracting the noise sources. The raw spectrum for one receive capture is shown in Fig.~\ref{fig:raw_receive_spectrum}.

\begin{figure}
    \centering
    \includegraphics{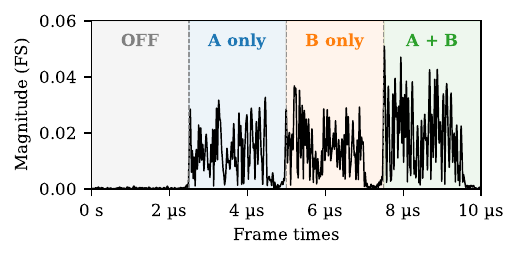} 
    \caption{Frame detection for an example receive capture. Filtered signal magnitude at nodes $N_0$, shifted using the detected frame offset. The shaded regions indicate the four source states.}
    \label{fig:frame_detection}
\end{figure}

\subsection{Frame Detection and State Extraction}
The transmitted sequence contained four states: both sources off,
source A only, source B only, and both sources active. Each state had
a \SI{2.5}{\micro\second} slot consisting of a \SI{2}{\micro\second} noise source and a \SI{500}{\nano\second} guard band. This corresponds to 500 samples per state giving a total of 2000 samples per waveform collected. Since the receivers and noise transmitters are not synchronized the resulting received frame does not necessarily begin at the start of the sequence. As a result, a recorded frame could begin within one source state and end within a later occurrence of that state. 
Due to this the frame shift was detected and corrected before extracting the different noise states. 

To detect the frame shift a reference waveform was generated using the same noise waveform, sample rate, bandwidth, digital frequency offset, and random seeds used during transmission. This reference was circularly cross-correlated with the received waveform to compare the known transmitted sequence with every possible sample offset. Conceptually, the reference and received waveforms were shifted relative to one another by one sample at a time, and the correlation magnitude was calculated at each lag. The corresponding lag that produced the largest correlation magnitude was selected as the estimated frame offset. This offset, together with the known waveform timing, was then used to identify the locations of the four source states. One note is that only complete source intervals were used and any interval split between the beginning and end of the next received record was discarded. The output of the correlation for one of the received frame and the corresponding corrected frame is shown in Fig.~\ref{fig:frame_detection}.

\subsection{Image Reconstruction}
Once each frame was corrected, the image was then reconstructed. First, each sliced source interval was used to calculate one complex visibility sample for its corresponding baseline. The extracted samples, $z_{\mathrm{rx},q,s,n}[m]$, from receiver $q$, state $s$, baseline $n$, and receiver capture $r$ calculated the cross-correlation
\[
\widehat{V}_{\mathrm{rx},s,n}[r]
=
\frac{1}{M}
\sum_{m=0}^{M-1}
z_{\mathrm{rx},0,s,n}[m]\,
z_{\mathrm{rx},1,s,n}^{*}[m]
\]
where $M$ is the total number of samples and $m$ is the sample index. The output gives the phase difference between the two receiver signals. The visibility samples from all received frames for each individual baseline was coherently averaged over all $r$ at each baseline before calculating their magnitude and phase. This produced four averaged complex visibility measurements at every baseline corresponding to the OFF, A-only, B-only, and A+B states.
The OFF-state visibility was kept as a background reference measurement. Next, these visibility were ordered were ordered by baseline, giving sixteen measurements from $1.5\lambda$ to $9\lambda$ in increments of $0.5\lambda$. For each state the reconstructed image was then created by taking the \ac{ift} of the sampled visibility, giving
\[
\widehat{I}_s(l)
=
\sum_{n=1}^{N_D}
\overline{V}_{s,n}
e^{j2\pi u_n l}
\]
where $N_D=16$ is the number of measured baselines, $u_n=D_n/\lambda$ is the corresponding spatial frequency coordinate, and $l=\sin(\theta)$ is the direction cosine.
The same reconstruction was performed separately for the A-only, B-only, and A+B measurements, producing three separate images. The visibility samples were zero padded to 1024 points before the \ac{ift}. All three images were then plotted with the x-axis as the angle and the y-axis as the power in decibel scale. The expected target directions of $-6.28^\circ$ and $12.95^\circ$ were indicated using red vertical lines.

\begin{figure}
    \centering
    \includegraphics{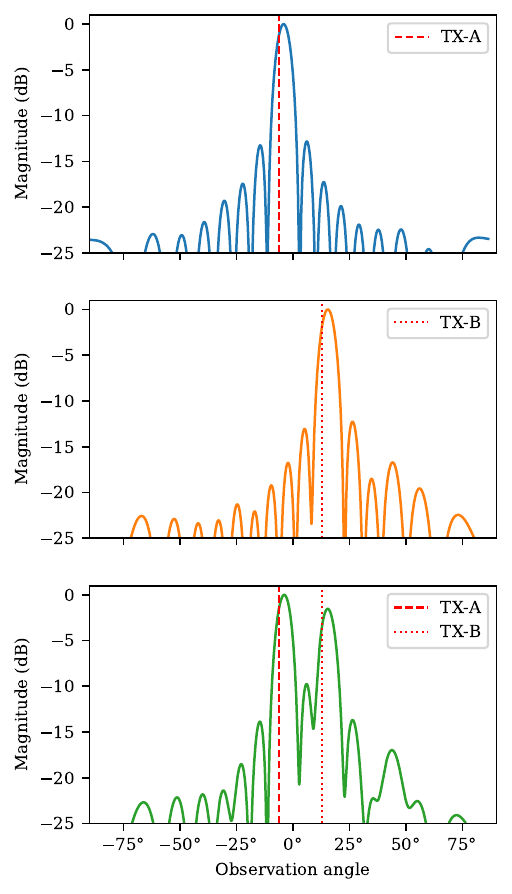}
    \caption{Measured one-dimensional Fourier-domain images for (top) source A only, (middle) source B only, and (bottom) both sources active. The images were reconstructed from sixteen baseline measurements ranging from $1.5\lambda$ to $9\lambda$ in $0.5\lambda$ increments. The red dashed lines indicate the expected source directions of $-6.28^\circ$ and $12.95^\circ$.}
    \label{fig:measured_imaging_results}
\end{figure}

\subsection{Imaging Results}

The three reconstructed images are shown in Fig.~\ref{fig:measured_imaging_results}. For the single source measurements, a clear response is seen near the expected direction of each source. When both sources were active the reconstructed image also contained two separate peaks corresponding to the two source directions. Therefore, the distributed receiver was able to reconstruct the individual sources as well as the complete two-source scene. The measured peak for source A was located at approximately
$-4.0^\circ$ in comparison with its expected direction of $-6.28^\circ$. This gives an  angular error of $\left|\Delta\theta_{\mathrm{A}}\right|\approx2.28^\circ$.
Similarly the measured peak for source B was located at $15.5^\circ$ compared with its ideal direction of $12.95^\circ$. The corresponding error was $\left|\Delta\theta_{\mathrm{B}}\right|\approx2.55^\circ$.
This residual angular error is within the margin of error for the placement of the transmit and receive antennas for the experiment.

\section{Conclusion}
A one-dimensional Fourier-domain imaging experiment using a fully-digital coordination distributed receiver was experimentally demonstrated. The system utilized two \ac{cots} Ettus X310 \acp{sdr} operating without a shared reference clock or external frequency standard. The expected source directions were $-6.28^\circ$ and $12.95^\circ$ and the reconstructed image peaks were reconstructed to within \ang{3} for a multi-target scene. 
This demonstrates that the coordination technique removed the unwanted timing, frequency and phase differences caused by the receivers independent clocks enabling the two source responses to clearly resolved and proves the feasibility the wirelessly coordinated interferometric imaging system. 


\bibliographystyle{IEEEtran}
\bibliography{reference}

\end{document}